\documentstyle[twocolumn,aps]{revtex}
\begin{document}
\draft
\title{ On an SO(5) unification attempt for the cuprates}
\author{G. Baskaran$^{\ddag}$ and P.W. Anderson}
\address{
Joseph Henry Laboratories of Physics\\ 
Princeton University\\ 
Princeton, NJ 08540 } 
\date{ \today }
\maketitle
\narrowtext

Zhang\cite{z} in his paper `A unified theory based on SO(5) 
symmetry of superconductivity and antiferromagnetism, Science 
$\bf 275$ 1089 (97)' has argued that there exists an approximate
global SO(5) symmetry in the low temperature phase of the 
high Tc cuprates.  This suggestion  contains 
a five component order parameter: three components correspond to a 
spin 1, charge zero particle-hole pair condensed at the center of 
mass momentum $(\pi, \pi)$, corresponding to antiferromagnetic 
order in the vicinity of the Mott insulating state; while the last 
two components correspond to a spin singlet, charge $\pm 2e$ 
cooper pair of orbital symmetry $d_{x^2 - y^2}$ condensed in the 
zero momentum state, corresponding to superconductivity in the
doped Mott insulator.  He identifies the operator 

\begin{equation}
\pi = \sum (cos k_x - cos k_y) c^{\dagger}_{{\bf k}\uparrow}
c^{\dagger}_{-\bf{k} +{\bf\pi} \uparrow}
\end{equation}
as one of the generators which rotates the 5-vector between the
spin and charge direction.  The doping density apparently acts
like a pseudo magnetic field for the order parameter.

For an exact SO(5) symmetry breaking scheme the generator $\pi$ should 
annihilate the vacuum (as it will be a zero momentum Goldstone
mode creation operator)  or its spectral function should have a
pole of finite residue at zero energy.  However, if the pole
has a finite but small energy, small compared to transition
temperatures, one can claim an approximate symmetry. 
Such a state above the top of the two-particle continuum
is referred to as an ``antibound'' state.
 Demler 
and Zhang\cite{dz} (DZ) claim that the spectral function of the 
$\pi$ operator has a pole around 41 meV for the cuprate YBCO based 
on a fermi
liquid t-matrix analysis, and identify the pole ({\em i.e.\/}
the antibound state) with a resonance
in neutron spectroscopy.

There are many criticisms which can be made of these ideas, from
mathematical ones to contradictions of the model with experimental
features of the cuprates, to deep general questions of physics.
The latter are more conclusive and less arguable, but perhaps 
initially it is essential to demonstrate the implausibility of 
this work so we shall start at the simplest and work up.
First let us list our objections:

1) (due to M. Greiter\cite{gr}) The Hubbard gap is ignored, the 
chemical potential misidentified and issues of projections are
not discussed.

2) The model Hamiltonian omits a large term , $t'$, which is 
experimentally observed and which displaces the resonance  and
possibly removes it.

3) The exchange terms in the Hamiltonian have little resemblance to
the correct ones, and the antibound state from magnetic 
interactions disappears once this error is corrected.

4) The conventional Hubbard Hamiltonian omits longer range coulomb
interactions, which are not important in the usual low energy physics
but are very much so here.  (We learned that Greiter also has brought
up this point in his second comment\cite{gr}).

5) The antiferromagnetic and superconducting phases each derive from a
more fundamental thermodynamic phase, the Mott insulator and the
metal respectively. The Mott insulator and the
metal can not be related to a quantum critical point, by Elitzur's
theorem, since they differ by a local gauge symmetry. These phases
hence have no locally stable, homogeneous intermediate phases, and
can not be deformed continuously into each other, certainly not by an
operator as simple as an SO(5) rotation.

We elaborate our points now.
1)  Greiter has brought up the importance of the upper Hubbard band
and the problems related to treating the chemical potential in a mean field
fashion without invoking projections.  We do not wish to go into the
Greiter - Demler, Zhang\cite{gr,dz1} discussion on the issue of 
chemical
potential.  The following is our comment on the projection issue.  The
spectral function of the $\pi$ operator will have a range from 
zero (the ground state) to an energy of the order of $U$.  The average 
energy of the state created by $\pi$ operator is unimportant.  
The issue is whether
the spectral function has a low energy pole with finite residue, and 
if so, whether the $\pi$ operator projected (renormalized) to this pole
continues to be a generator obeying the old SO(5) algebra or not.
The projected operator may be defined as

\begin{equation}
\pi_\alpha = P_\alpha \pi P_\alpha
\end{equation}
where $P_\alpha = |\alpha><\alpha| $ is the projection operator; and
the state $|\alpha>$ is the many body (N + 2)-particle state 
corresponding to the pole.  Projections
to subspaces can change the commutation relations.  For example,
in the case of the Mott 
insulator, the $\pi$ operator projected to the lower Hubbard band 
creates  
only spin and hence it is incapable of rotating between the charge and 
spin sectors of SO(5). 

2) The known fermi surfaces of cuprates are inconsistent with 
the bipartite model used by DZ,  which contains only nearest
neighbor hopping.  The correct form of dispersion, as has also been
noted by DZ, is

\begin{equation}
 \epsilon_{k} = - 2t (cos k_x + cos k_y) - 2t' ( cos(k_x + k_y) + 
(cos(k_x - k_y) )
\end{equation}
\noindent
with $t'$ of the order of $1\over2$ $t$. The $t'$ term does not commute with
the operator $\pi$ and spreads its spectrum\cite{dz4}
 over 0.1 eV. Any resonance
must be outside this range, presumably an antibound state above it.
It is also easy to show that the spreading of the two particle spectrum
at$(\pi,\pi)$ reduces the already small antibinding energy\cite{dz2} 
$(0.1 J \approx 10$ to $15 meV)$.
This is because the spin triplet antibinding state  
has a p-symmetry in the relative orbital co-ordinate forced by 
antisymmetry,  and a p-wave
bound state\cite{dz3} (to be precise $p_x - p_y$ symmetry) does not 
effectively make use of short range attraction in view of the node 
at the origin.  

That spreading of the two particle band (corresponding to relative
motion)  will destroy the two particle
antibound state can be already inferred from DZ's results directly.
A change of the center of mass momentum by only  about $\pi \over {50}$ 
is enough to remove the antibinding energy (figure 1 in DZ). This 
happens when the two particle band width is $\approx 0.1 J$,
comparable to the antibinding energy at $(\pi,\pi)$.  
This is a consequence of the fact
that in a tight binding situation nearest neighbor attraction will
give a p-wave binding only when its strength is of the order of 
the band width.

3) The Hamiltonian (equation 1 of DZ) used by DZ includes a nearest 
neighbor exchange term in a Hubbard model.  This is, rather 
confusingly, included in
order to capture the higher order effects in $1\over U$. 
The derivation of this exchange term (see PWA \cite{pwa1}) 
leads to the expression 

\begin{equation}
J_{ij} ({\bf S}_i . {\bf S}_j - {1\over4} n_i n_j )
\end{equation}
not $~~~ {\bf S}_i . {\bf S}_j ~~~$ alone:  hence there is no interaction 
(repulsion) between parallel spins at all.  This means\cite{dz5}
the absence of an antibound state to leading order in J ;({\em i.e.\/}
absence of 41 meV resonance for YBCO within DZ's scheme.  
In fact, the correct 
Hamiltonian would also contain a ferromagnetic true exchange
term (whose origin is potential rather than kinetic), and the net 
effective interaction between parallel spins on
nearest neighbor sites would be attractive, thereby removing any
antibinding tendency by magnetic interactions.  

We also wish to point out that any induced magnetic exchange
interaction between two electron spins will have the form, 
\begin{equation}
J_{ij;kl} \sum_\sigma c^\dagger_{i\sigma}c_{j\sigma}\sum_{\sigma'}
c^\dagger_{k\sigma'}c_{l\sigma'}
\end{equation}
in both 
metallic and insulating states, irrespective of the value of U,  
in view of the kinetic origin of the induced magnetic exchange
interactions in tight binding systems like cuprates.  When
we specialize $J_{ij;kl}$ to two sites we recover equation
(4).  

4) While considering excited states, interactions that were not
very effective at low energies may start playing a major role. 
There is a strong repulsive term which prevents a parallel spin 
two particle resonance at low energies: the unscreened part of the 
coulomb repulsion between pairs of particles on neighboring sites.  
This must be of the order of an eV.  It is easy to show using 
the same t-matrix analysis that this term brings back the 
antibound state but at about 1000 meV and not 40 meV !  In most
Hubbard model calculations this term is neglected, or taken 
care of through Hartree like approximations, because `$U$' is
so much larger, but of course it is always present and where
one is discussing a resonance in the particle-particle channel
it must be taken into account as a final state interaction.

Incidentally, these antibound states share a common origin
with the two particle anti bound state in the singlet channel whose 
energy is of the order of onsite $U$; in the triplet channel 
it is of the order of the nearest neighbor $V$ which is the major 
repulsion felt by parallel spins.

5) The presence of all of the above terms means simply that the 
delicate balance created by the artificially restricted model used
in the SO(5) theory is, indeed, an artifact of model building and 
has little relation to physical reality.  But there is a much
deeper and more profound difficulty with theories of the 
cuprates based upon a quantum critical point.  Of these, the most
ambitious is the SO(5) hypothesis of Zhang , but others have
been suggested (e.g. by Sachdev et al.\cite{qcp})

The general idea is to propose that thermodynamic phases are
characterized by an order parameter such as superconducting
pair wave functions or an (anti)ferromagnetic moment.  One 
imagines a critical point at T = 0, at which this order parameter 
changes continuously from one type to another (or ceases to 
exist)  as some parameter -
in the case of cuprates, the doping - is varied.  One has then a
multi-dimensional space of some sort, containing an order parameter
vector of finite dimensionality and this vector rotates from 
one sector of this space to another at the critical point.  The 
associated T = 0 quantum critical fluctuations spread its influence 
into a range of finite T and doping.

This picture does not take into account the basic physics of the 
two states.  The antiferromagnet is a Mott insulator, and it is an 
antiferromagnet because it is a Mott insulator, not vice versa.
Superexchange is a consequence of the insulating state (see
PWA\cite{pwa2}).  The insulating state has a 
large gap for charge fluctuations, and the antiferromagnetic order 
parameter forms in the subspace formed by bare spins.

The superconductor, on the other hand, forms in a metal, and is
a property of low-energy quasiparticle-like excitations near
a fermi surface.  The fermi surface is meaningless in the insulator,
but is crucial to the superconductor.

The dominant transition, then, is between metal and insulator,
each being states described by an infinite dimensional order
parameter\cite{gb}.  The transition 
between these two states can not be continuous, and is not; the
insulator has a discontinuity in chemical potential, 
$ \mu = {{\delta F}\over{\delta  n }}$and is therefore a cusp 
of the free energy.
States of intermediate doping cannot be homogeneous and are not.
When doping is done by mobile charges (laser doping, or oxygen
doping in $La_2 Cu O_{4+\delta}$) the system segregates.  
In summary, the order parameter characterizing antiferromagnetism
and superconductivity relate to subspaces of Hilbert space
which are algebraically inequivalent and can not be mapped on
each other;  thus neither can the order parameters themselves.
There is no underlying quantum critical point expressing an 
essential continuity between the two phases.

\section {Acknowledgement}

We would like to acknowledge Steve Strong for bringing to our
attention the preprint by Martin Greiter and Steve Strong
and David Clarke for discussions.  This work was supported 
by the National Science Foundation Grant DMR - 9104873.


\begin{references}
\bibitem{dz2} DZ get the energy  of the $\pi$ resonance $E_{\pi}$
by adding this antibinding energy $E_b \approx 0.1J$  to the center 
of the two particle $(\pi,\pi)$ band which is assumed to be around 0.2J 
above the ground state; $E_{\pi} \approx 2\mu + E_b = 0.2 J + 0.1J
\approx 40 meV$.
\bibitem{dz4} Pauli blocking of the two particle states reduce this
spread.  However, a spread as little as 10 meV seems to be
enough to remove the antibound state.
\end{references}
\end{document}